\documentclass{iucr}
\RequirePackage{graphicx}
\usepackage{graphicx}

     \paperprodcode{a000000}      % Replace with production code if known
     \paperref{xx9999}            % Replace xx9999 with reference code if known
     \papertype{CP}               % Indicate type of article

     \paperlang{english}          % Can be english, french, german or russian

     \journalcode{J}              % Indicate the journal to which submitted
     \journalyr{2010}
     \journaliss{1}
     \journalvol{43}
     \journalfirstpage{000}
     \journallastpage{000}
     \journalreceived{0 XXXXXXX 0000}
     \journalaccepted{0 XXXXXXX 0000}
     \journalonline{0 XXXXXXX 0000}

\begin{document}                  % DO NOT DELETE THIS LINE

\title{Fast computing of scattering maps of nanostructures using graphical processing units}
\shorttitle{GPU-accelerated computing of scattering maps for nanostructures}

\cauthor[a,b]{Vincent}{Favre-Nicolin}{Vincent.Favre-Nicolin@cea.fr}{}
\author[c]{Johann}{Coraux}
\author[d]{Marie-Ingrid}{Richard}
\author[e]{Hubert}{Renevier}

\aff[a]{CEA-UJF, INAC, SP2M, Grenoble, \country{France}}
\aff[b]{Universit\'e Joseph Fourier, Grenoble, \country{France}}
\aff[c]{Institut N\'eel, CNRS-UJF, Grenoble, \country{France}}
\aff[d]{Universit\'e Aix-Marseille, IM2NP, UMR CNRS 6242, FST, Marseille \country{France}}
\aff[e]{Laboratoire des Mat\'eriaux et du G\'enie Physique, Grenoble INP—MINATEC, Grenoble,\country{France}}


\maketitle

\begin{synopsis}
Scattering calculations from atomistic models using Graphical Processing Units are presented, and compared to the speed achieved using normal CPU calculations. An open-source software toolbox (PyNX) is presented, with a few examples showing the fast calculation of scattering maps from strained nanostructures, including grazing-incidence conditions.
\end{synopsis}

\begin{abstract}
Scattering maps from strained or disordered nano-structures around a Bragg reflection can either be computed quickly using approximations and a (Fast) Fourier transform, or using individual atomic positions. In this article we show that it is possible to compute up to $4.10^{10}$ $reflections\cdot atoms\cdot s^{-1}$ using a single graphics card, and we evaluate how this speed depends on number of atoms and points in reciprocal space. An open-source software library (PyNX) allowing easy scattering computations (including grazing incidence conditions) in the Python language is described, with examples of scattering from non-ideal nanostructures.
\end{abstract}

\section{Introduction}

Efficient computing of X-ray (and neutron) scattering from crystals has been the subject of intense work since computers became available. Except in the case of small structures ($<$1000 atoms) or small number of reflections ($<$1000), the method of choice has long been to use the Fast-Fourier Transform \cite{ten_eyck_crystallographic_1973,immirzi_general_1973,ten_eyck_efficient_1977} of the crystal's scattering density. By computing this density inside the crystal's unit cell over a suitable grid \cite{ten_eyck_efficient_1977,langs_elimination_2002}, it is possible to compute structure factors at nodes of the reciprocal lattice.

In the case of strained \cite{takagi_dynamical_1969} or disordered \cite{butler_calculation_1992} crystals, the scattering must take into account a large part of the crystal (or possibly the entire crystal) instead of a single unit cell, in order to describe the departure from an infinite, triperiodic structure. This requires either using approximations, or a large computing power. Moreover, both strain and disorder lead to non-discrete scattering, so that the scattered amplitude must be evaluated on a fine grid around or between Bragg diffraction peaks. This type of computations can greatly benefit from fast calculations, which we will present in this paper.

This article is organized as follows: in section \ref{secScattering} we describe the formulas used for computing the scattering from an atomistic model, how it can be efficiently computed using a Graphical Processing Unit (GPU), and what performance can be achieved. In section \ref{secPyNX} we present the open-source package PyNX which can be used to easily compute scattering with little programming knowledge. In section \ref{secApplication} a few examples are given.

\section{Scattering computing from an atomistic model using a GPU}\label{secScattering}
\subsection{Theory}
X-ray and neutron scattering can generally be calculated, in the kinematic approximation, as:
\begin{equation}\label{eqn:eqnRhoFT} 
A(\mathbf{S})=\int_V \rho(\mathbf{r})\exp(2i\pi\mathbf{S}\cdot \mathbf{r}) dV = FT\left[\rho(\mathbf{r})\right]
\end{equation}
where $A(\mathbf{S})$ is the scattered amplitude, $\textbf{S}$ is the scattering vector, $\rho(\textbf{r})$ represents the scattering density (electronic or nuclear) at position $\textbf{r}$ inside the crystal, and $FT$ denotes the Fourier transform. This equation can be used to determine the scattering from a crystal as long as $\rho(\textbf{r})$ is described on a grid fine enough to resolve the atomic positions, which is easy if the crystal can be described from a single unit cell.

In the case of an aperiodic object (crystal with an inhomogeneous strain or disordered), it is simpler to compute the scattering from an atomistic model, which can be obtained using reverse Monte-Carlo \cite{nield_interpretation_1995,proffen_analysis_1997,welberry_analysis_1998} , atomic potentials combined with molecular dynamics or a direct minimization of the crystal energy \cite{keating_effect_1966,stillinger_computer_1985,tersoff_new_1988,plimpton_fast_1995,niquet_electronic_2006,katcho_n_a_structural_2009}. The scattered amplitude is then derived from the atomic positions:

\begin{equation}\label{eqn:eqnScattAtoms} 
A(\textbf{S})=\sum_{j} f_j(S) \exp(2i\pi\mathbf{S}\cdot \mathbf{r}_j )
\end{equation}
where $f_j(S)$ is the scattering length (either the Thomson scattering factor for X-rays or the nuclear scattering length for neutrons) of atom $j$ and $\textbf{r}_j$ its position.

The number of floating-point operations ($flop$) required to evaluate equation (\ref{eqn:eqnScattAtoms}) is approximately equal (see section \ref{secImplemGPU}) to:
\begin{equation}
N_{flop}\approx 8\times N_{atoms}\times N_{reflections}
\end{equation}

For a structure with $N_{atoms}=8\times10^6$ (e.g. a cube of silicon of $54\times54\times54\ nm^3$) and $N_{refl}=256\times256$ points in reciprocal space, this corresponds to $4.2\times10^{12}\ flop$, which can be compared to the current computing power of today's consumer micro-processors of $5-10\ \times 10^9 flop\cdot s^{-1}$ per core.

In the case of large nano (and micro)-structures, for which the description of the atomic structure is not possible in practice, a model based on continuum elasticity can be used, either with an analytical or numerical approach (see \citeasnoun{stangl_structural_2004} and references within). The most popular method in the case of epitaxial nanostructures currently is the Finite Element Method- see \citeasnoun{wintersberger_algorithms_2010} for a recent discussion. This model can then be used to calculate the scattering using groups of atoms:
\begin{equation}
\label{eqn:eqnScattBlock} 
A(\textbf{S})=\sum_{j} F_j(\textbf{S}) \exp(2i\pi\mathbf{S}\cdot (\mathbf{r}_j^0+\textbf{u}_j)) 
\end{equation}
where $F_j(S)$ is the structure factor for the $j^{th}$ block of atoms (generally a group of unit cells), $\textbf{r}_j^0$ its original position, and $\textbf{u}_j$ its displacement from the ideal structure.

Assuming that all blocks of atoms are identical and on a triperiodic grid, it is possible to rewrite Eqn. \ref{eqn:eqnScattBlock} as:
\begin{equation}
\label{eqn:eqnScattBlockFT} 
A(\textbf{S})\approx\ F(\textbf{S})\ FT\left[  \exp(2i\pi\mathbf{H}\cdot \textbf{u}_j) \right]
\end{equation}
where $\textbf{H}$ denotes the Bragg peak position around which the calculation is made, $F(\textbf{S})$ is the structure factor calculated for a block of atoms, and $\textbf{u}(\textbf{r})$ the displacement field inside the crystal.

If the composition of the blocks of atoms vary (e.g. due to interdiffusion), it is also possible to include a variation of the average scattering density in the $FT$ \cite{takagi_dynamical_1969}:
\begin{equation}\label{eqn:eqnRhoFTDispl} 
A(\mathbf{S})\approx\ F(\textbf{S})\  FT\left[\overline{\rho}(\mathbf{r})\ \exp(2i\pi\textbf{H}\cdot\textbf{u}(\textbf{r}))\right]
\end{equation}

where $\overline{\rho}(\mathbf{r})$ is the relative scattering density in the crystal.

Both equation (\ref{eqn:eqnScattBlockFT}) and (\ref{eqn:eqnRhoFTDispl}) allow the use of a \textit{fast} Fourier transform, but are only valid as long as:\footnote{A more detailed study of this approximation will be presented in another article devoted to X-ray coherent diffraction imaging in Bragg condition}
\begin{equation}
\mid(\textbf{S}-\textbf{H})\cdot \textbf{u}(\textbf{r})\mid\ll 1
\end{equation}

Moreover, use of equations (\ref{eqn:eqnScattBlockFT}) and (\ref{eqn:eqnRhoFTDispl}) with a \textit{fast} Fourier transform restricts the computation of scattering on a triperiodic grid in reciprocal space - this is a limitation since modern data collection often use 2D detectors, and the measured points in reciprocal space are located on a \textit{curved} surface (the projection of the detector on Ewald's sphere). Furthermore, as the resolution in reciprocal space is inversely proportional to the size in real space, analysis of high-resolution data using a FFT calculation demands a large model - even if the extent in reciprocal space is very limited.

Therefore, even if the speed of the FFT is optimal for large crystalline structures - for $N$ points in real space, $N$ points in reciprocal space are calculated with a cost proportional to $N\cdot \log(N)$ instead of $N^2$ - it is still interesting to consider a \textit{direct} computation using equation (\ref{eqn:eqnScattAtoms}) or (\ref{eqn:eqnScattBlock}) because it allows computation for:
\begin{itemize}
\item \textit{any} assembly of points in reciprocal space
\item from \textit{any} structural model (no matter how severely distorted or disordered) 
\end{itemize}

\subsection{Implementation on a GPU}\label{secImplemGPU}

In order to achieve the calculations in a reasonable time, it is useful to consider current graphics cards as general-purpose GPU. This has already been reported in the scope of crystallography, for computing scattering maps from disordered crystals \cite{gutmann_accelerated_2010}, powder pattern computing using the Debye equation \cite{gelisio_real-space_2010}, and for single-particle electron microscopy \cite{schmeisser_parallel_2009}.

To summarize basic principles behind GPU computing, it is possible to accelerate any calculation provided that:
\begin{enumerate}
\item it is \textbf{highly parallel}, \textit{i.e.} the same formula must be applied on large amounts of data, independently
\item the number of \textbf{memory} transfers required is much smaller than the number of mathematical operations
\item the calculation pathway is determined in advance (at compilation time), which excludes any \textit{if...then...else} operation in the inner computation loop
\end{enumerate}

Moreover, many classical functions (e.g.: $exp$, $log$, $sqrt$, fused $sin-cos$ evaluation,...) are highly optimized on GPUs - an algorithm requiring many such operations will be greatly accelerated.

Equation (\ref{eqn:eqnScattAtoms}) fulfills all requirements, assuming that both the number of atoms and the number of points in reciprocal space are large ($\gg$1000).

\begin{figure}\label{figSpeed}
\includegraphics[width=0.95\textwidth]{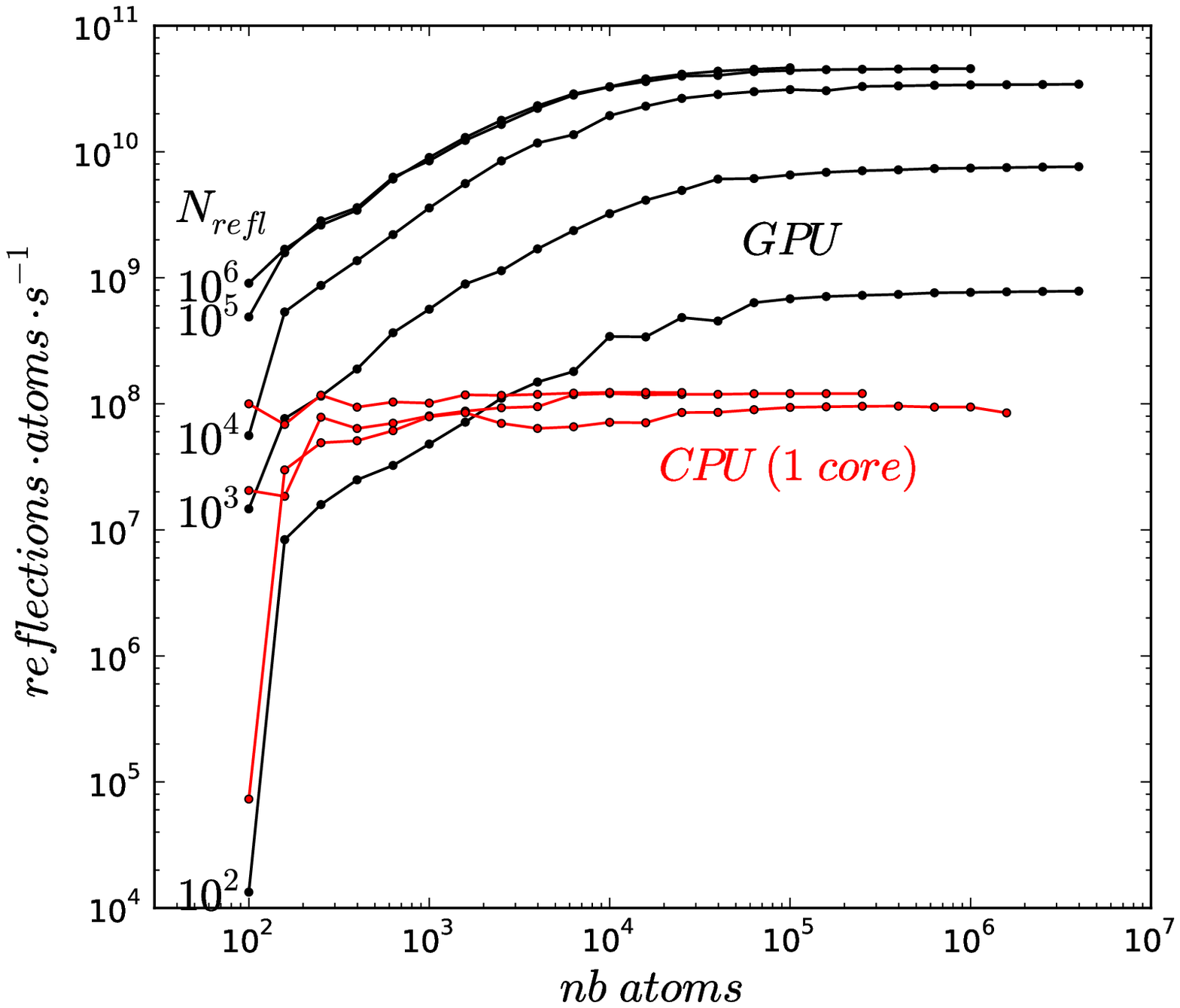}
\caption{Computing speed depending on the number of atoms and reflections. These tests were run on a single nVidia GTX295 graphics card, using in parallel the two multiprocessors available on the card. $N_{refl}$ indicates the number of reflections for the GPU calculations (black lines). The CPU (Central Processing Unit) curves (red lines) correspond to a computing using a vectorized (SSE-optimized) C++ code running on a \textit{single} core of an Intel Core2 Quad Q9550 running at 2.83 GHz, for $N_{refl}=10^2,\ 10^3, 10^4$ (the curves for $10^3$ and $10^4$ are almost identical).}
\end{figure}

The implementation presented in this article uses the CUDA \cite{nvidia_cuda_2010} toolkit. It is beyond the scope of this article to detail the exact algorithm used for computation, as the implementation is freely available as an open-source project (see section \ref{secPyNX}). However, it should be noted that the calculations are made in parallel for all reflections, with the atomic coordinates shared between parallel threads (to minimize memory transfers) - this method is optimal for large number of atoms. For some configurations (large number of reflections and small number of atoms), it may be more optimal to parallelize on the atoms and share the reflection coordinates between parallel processes.

The achieved speed is shown in Fig. \ref{figSpeed}, for a calculation of scattering for a random list of points in reciprocal space, and random coordinates for the atoms - the occupancy of all atoms is assumed in this test to be equal to 1, and the atomic scattering factor is not evaluated- in practice the atomic scattering factors can be factorized and represent a negligible amount of computing  (see section \ref{sec:SecExampleInAs}) - the same is true for Debye-Waller factors.

As can be seen in Fig. \ref{figSpeed}, there is a strong dependence of the speed with the number of reflections and the number of atoms per second - the maximum speed ($\approx 46\times 10^9\ reflection\cdot atoms\cdot s^{-1}$) is only reached if both numbers are larger than $10^4$. 

Each couple (reflection, atom) corresponds to 8 floating-point operations (3 multiplications, 4 additions, one $sincos$ evaluation)\footnote{Note that although the $sincos$ operation is hardware-accelerated, it is 4 to 8 times slower than a simple addition. If the $sincos$ evaluation is counted as 4 $flop$, the achieved speed is $\approx$500 $Gflop\cdot s^{-1}$.}, so that the overall speed is equal to $\approx$367 $Gflop\cdot s^{-1}$. This can be compared to the peak theoretical speed of 1.7 $Tflop\cdot s^{-1}$ for this graphics card, which is only achieved when using fused add-multiply operations, without any bottleneck due to memory transfers.

By comparison, when computing on the CPU (see Fig. \ref{figSpeed}), the maximum speed is reached sooner: for 100 atoms and $>$1000 reflections, or for $>$1000 atoms when using $>$100 reflections. The top speed for a \textit{single} core (Intel Core2 Quad Q9550 running at 2.83 GHz), using SSE-vectorized\footnote{SSE: Streaming SIMD Extensions ; SIMD: Single Instruction, Multiple Data} sine and cosine functions \cite{pommier_julien_simple_2008}, is $122\times10^6\  reflection\cdot atoms\cdot s^{-1}$ - $\approx$380 times slower than the GPU version. An \textit{un}-optimized (without using SSE code) C++ code runs about 3 times slower, or $\approx$1000 times slower than the GPU version. Using multiple cores, the speed increases linearly (except for small number of atoms) with the number of cores.

\subsection{Accuracy}
As was already pointed out by \citeasnoun{gutmann_accelerated_2010}, accuracy is an important issue since GPUs are most efficient when using single precision floating-point operations. Moreover, the accuracy of operations can be slightly relaxed \cite{nvidia_cuda_2010} compared to IEEE standards \cite{ieee_ieee_2008}. 

For example, since single-precision floating-point use 24 bits mantissa (\textit{i.e.} a relative accuracy of $\approx 10^{-7}$), precision may be expected to become problematic when the total scattered amplitude varies on more than 7 orders of magnitude.

\begin{figure}
\label{figAccuracy}\caption{Comparison of the accuracy of the computation of the scattering between single-precision floating-point and an analytical model. The calculation is made for a linear chain of atoms with (a) $10^3$ (b) $10^4$ and (c) $10^5$ atoms, using a perfectly periodic chain calculated using the GPU (black line), the analytical model (red line), and for a chain where the atoms are randomly displaced with a Gaussian distribution with a standard deviation of $10^{-3}$ (gray line). The atoms where located at $0,1,...,(N_{atom}-1)$, and the H coordinates are located every 0.001 in reciprocal lattice units (r.l.u.).}
\includegraphics[width=0.95\textwidth]{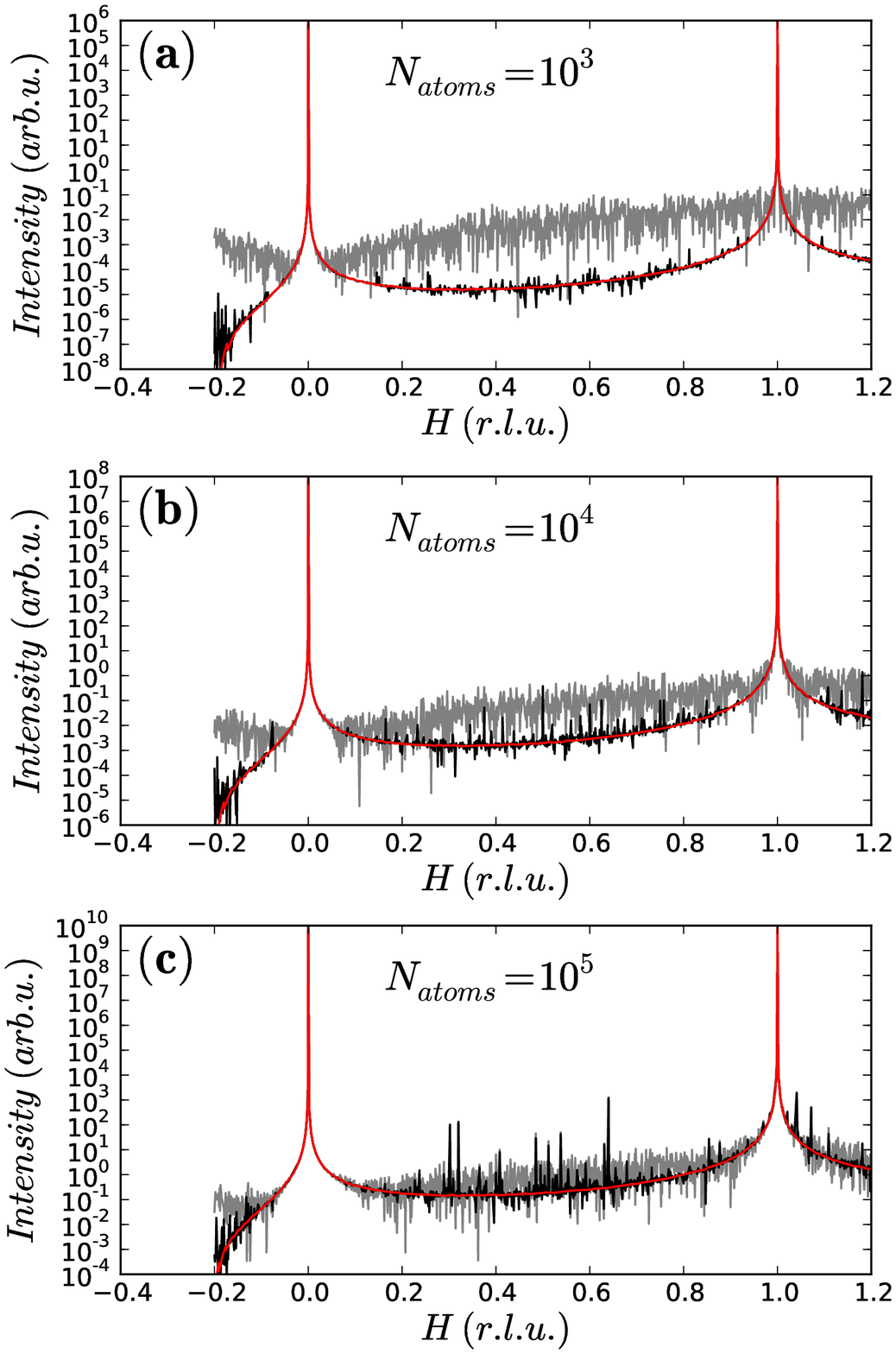}
\end{figure}

A simple test can be made: computing the scattering for a linear, perfectly periodic chain of identical atoms, and comparing it to the analytical formula: $A(h)=\frac{\sin(\pi N\times H)}{\sin(\pi H)}$ (where $H$ is the reciprocal lattice unit and $N$ the number of atoms in the chain). This is shown in Fig.\ref{figAccuracy}, for chains of atoms of different lengths. The discrepancy between the analytical calculation and the single-precision GPU calculation is clearly visible in the regions where the intensity  is minimal - however in practice, the dynamic range where the calculation is reliable is always larger than $10^8$, and most of the time ($>99\%$ of the points) around $10^{10}$ - these numbers refer to the intensity (squared modulus of the scattered amplitude).

Such a dynamic range should be sufficient for most applications, as the practical range for experimentally measured intensities is usually lower, except in the case of perfect crystals. In Fig.\ref{figAccuracy} a gray curve is superposed to the simulations, and corresponds to the GPU calculation for a chain of atoms with random displacements with a Gaussian standard deviation of $10^{-3}$ of their fractional coordinate. The error due to the single-precision computing is generally lower than the 'noise' level represented by the gray curve.

We have found that errors due to single precision floating point  calculations were not significant \textit{in practice}: indeed, most of the time structural models for which this type of computation is used are not ideal (see examples of simulated calculations using our code in \citeasnoun{tardif_strain_2010} and \citeasnoun{favre-nicolin_analysis_2010}) and therefore do not present a very large dynamic range (larger than 8 orders of magnitude).

It should however be noted that GPUs can also perform calculations using double precision floating point, but with a lower performance, as the number of available processing units are generally smaller (8 times in the case of CUDA graphics cards with capability less than 2.0 \cite{nvidia_cuda_2010}) than for single precision calculations. More recent graphics cards (available since mid-2010), using the Fermi architecture (\verb!http://www.nvidia.com/object/fermi_architecture.html!) provide a higher computing power dedicated to double-precision computing (about half the speed of single-precision).

\section{Open-source implementation: PyNX }\label{secPyNX}

Writing programs using GPU computing is a relatively complex process, as it is necessary to fine-tune the algorithm, notably in order to optimize memory transfers - which can make a very significant difference in terms of performance. For example an early version of the presented algorithm did not perform in a synchronized way between parallel computing threads, and its performance was $10\times$ slower than the final algorithm used. Moreover, all data has to be allocated both in the computer's main memory as well as on the graphics card, which can be tedious to write.

For this reason, we have written an open-source library, PyNX "Python tools for Nanostructure Xtallography", using the Python language. The main features of this software package are the following:
\begin{itemize}
\item Computing of scattering for a given list of atomic positions and points in reciprocal space does not require \textit{any} GPU-computing knowledge
\item it is possible to input either a list of $(x,y,z)$ coordinates, or also include their occupancy $(x,y,z,occ)$
\item the shape and order of the $(x,y,z)$ and $(h,k,l)$ coordinates (i.e. 1D, 2D or 3D, sorted or not) is irrelevant - all calculations are made \textit{in fine} on 1D vectors
\item the computation can be distributed on several GPUs - e.g. such cards as NVIDIA's GTX 295 are seen as two independent GPU units - the calculation is distributed transparently over the two GPU
\item a pure SSE-optimized CPU computation is also available when no GPU is available, and can take advantage of all the computing cores available.
\end{itemize}

Three modules are available: 
\begin{itemize}
\item \verb!pynx.gpu!, which is the main module allowing fast, parallel computation of $\sum_{j}\exp(2i\pi\mathbf{S}\cdot \mathbf{r}_j )$, either using a GPU or the CPU
\item \verb!pynx.fthomson!, which gives access to the analytic approximation for the X-ray atomic scattering factors \cite{prince_international_2004} extracted from the cctbx library \cite{grosse-kunstleve_computational_2002}
\item \verb!pynx.gid!, which provides transmission and reflection coefficients at an interface \cite{dosch_critical_1992}, which is required for grazing incidence diffraction analysis \cite{kegel_nanometer-scale_2000,schmidbauer_effects_2005,richard_multiple_2009} using the Distorted Wave Born Approximation (DWBA), as is demonstrated in section \ref{sec:DWBA}.
\end{itemize}

\section{Examples}\label{secApplication}

\subsection{Simple monoatomic cubic structure with $100\times 100\times 100$ unit cells}

\begin{figure}
\label{figStrainedCube}\caption{Example calculation of the 2D scattering of a strained cubic nanostructure (see text for details), around reflection $(004)$. Intensities are color-coded on a logarithmic scale.}
\includegraphics[width=0.95\textwidth]{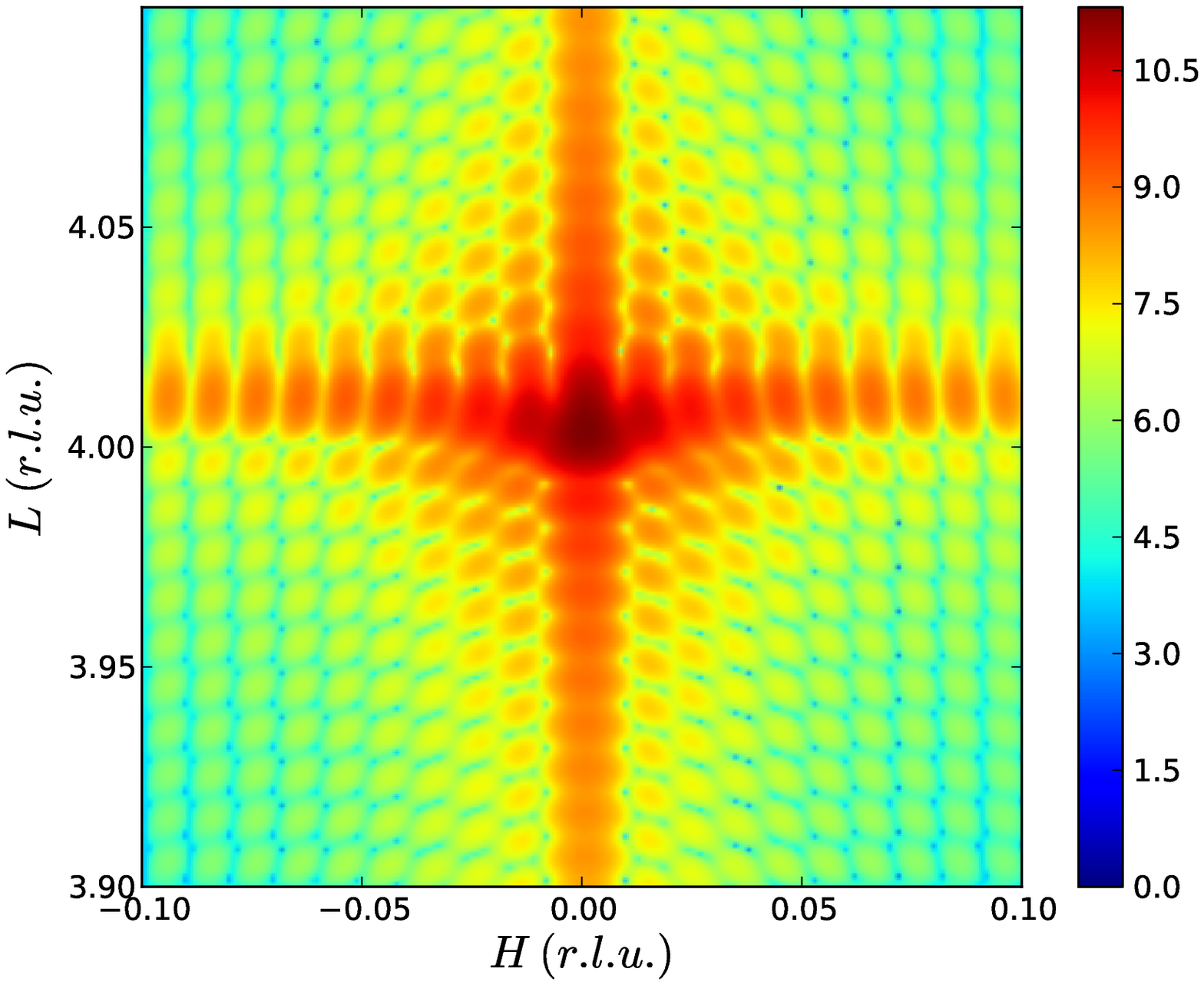}
\end{figure}

To compute the scattering around the $(004)$ reflection of a simple cubic structure with a lateral size of 100 unit cells, the following code is used:
\begin{verbatim}
#Import libraries
from numpy import arange,float32,newaxis,log10,abs
from pynx import gpu

#Create array of 3D coordinates, 100x100x100 cells
x=arange(-50,50,dtype=float32)
y=arange(-50,50,dtype=float32)[:,newaxis]
z=arange(0,100,dtype=float32)[:,newaxis,newaxis]

#HKL coordinates as a 2D array
h=arange(-.1,.1,0.001)
k=0
l=arange(3.9,4.1,0.001)[:,newaxis]

#The actual computation
fhkl,dt=gpu.Fhkl_thread(h,k,l,x,y,z,gpu_name="295")

#Display using matplotlib
from pylab import imshow
imshow(log10(abs(fhkl)**2),vmin=0,
   extent=(-.1,.1,3.9,4.1))
\end{verbatim}

In this example, the calculation takes 0.93s on a GTX 295 graphics card. The library used for graphics display is matplotlib (http://matplotlib.sourceforge.net/).

Of course scattering from this cube could be calculated analytically - if we introduce a simple displacement field in the z-direction: $u_z=10^{-6}\times z\times(x^2+y^2)$, the following line can be inserted after the "\verb! z=arange...!" instruction:
\begin{verbatim}
z=z+1e-6*z*(x**2+y**2)
\end{verbatim}

The computed diffraction map is shown in Fig.\ref{figStrainedCube}.

\subsection{Bi-atomic structure from file}\label{sec:SecExampleInAs}

In the previous example, the atomic scattering factor is not taken into account - since this factor is the same for all atoms of the same type, it is easy to group all atoms of the same type together and calculate first $\sum_{j}\exp(2i\pi\mathbf{S}\cdot \mathbf{r}_j )$, and then multiply it by the value of the atomic scattering factor dependent on the position in reciprocal space (and the energy if anomalous scattering terms are to be taken into account), as well as the Debye-Waller factor. These atomic scattering factors can be extracted from the \verb!pynx.fthomson! module.

Let us consider an InAs nano-structure, for which we have atomic coordinates in separate files \verb!In.dat! and \verb!As.dat!, each file having 3 columns corresponding to the x,y,z orthonormal coordinates (in nanometers). The scattering around reflection $(004)$ for this data can be calculated in the following way (the f' and f" resonant terms were taken manually from the cctbx library \cite{henke_low-energy_1982,grosse-kunstleve_computational_2002}):

\begin{verbatim}
#Import libraries
from numpy import arange,newaxis,sqrt,abs,loadtxt
from pynx import gpu,fthomson

#HKL coordinates as a 2D array
h=arange(-.1,.1,0.001)
k=0
l=arange(3.9,4.1,0.001)[:,newaxis]

#Load orthonormal coordinates
xAs,yAs,zAs=loadtxt("As.dat",unpack=True)
xIn,yIn,zIn=loadtxt("In.dat",unpack=True)

#Convert to fractional coordinates
xAs/=.6036
yAs/=.6036
zAs/=.6036
xIn/=.6036
yIn/=.6036
zIn/=.6036

#Compute scattering
fhklIn,dt=gpu.Fhkl_thread(h,k,l,xIn,yIn,zIn,
                          gpu_name="295")
fhklAs,dt=gpu.Fhkl_thread(h,k,l,xAs,yAs,zAs,
                          gpu_name="295")

#Apply scattering factors at 10keV
s=6.036/sqrt(h**2+k**2+l**2) 
fAs=fthomson.FThomson(s,"As")-1.64+0.70j
fIn=fthomson.FThomson(s,"In")+0.09+3.47j

#Full structure factor
fhkl=fhklAs*fAs + fhklIn*fIn

\end{verbatim}

%\subsection{Other examples: grazing-incidence diffraction}

%More complex examples (but beyond the scope of this article) are available on the PyNX website. Most notably, a specific module (\verb!pynx.gid!) is available for grazing incidence diffraction. This module allows to compute the complex refraction index of a crystalline material (the substrate) and determine the reflection and transmission coefficients at the interface. It is therefore possible to simulate grazing incidence X-ray scattering using the DWBA approximation.

\subsection{Grazing-Incidence Diffraction using the DWBA approximation}
\label{sec:DWBA}

A specific module (\verb!pynx.gid!) is available for grazing incidence diffraction - this module allows to compute the complex refraction index of a crystalline material (the substrate) and determine the reflection and transmission coefficients at the interface \cite{dosch_critical_1992}. It is therefore possible to simulate grazing incidence X-ray scattering using the DWBA approximation, by taking into account the reflections before and after diffraction by the object at the surface, which influence the shape of the scattering in reciprocal space \cite{kegel_nanometer-scale_2000,schmidbauer_effects_2005,richard_multiple_2009}.

\begin{figure}
\label{figDWBA}\caption{Simulated scattering from a germanium quantum dot on a silicon substrate, using the DWBA approximation. The calculation is made around the in-plane $(400)$ reflection, and is plotted against $H$ (reciprocal lattice unit) and the outgoing angle $\alpha_f$. The location (outgoing angle) of the intensity maximum as a function of $H$ varies as the in-plane strain changes with the height of the corresponding layer in the dot, due to interferences between the four scattered beams \cite{kegel_nanometer-scale_2000}. See text for details.}
\includegraphics[width=0.95\textwidth]{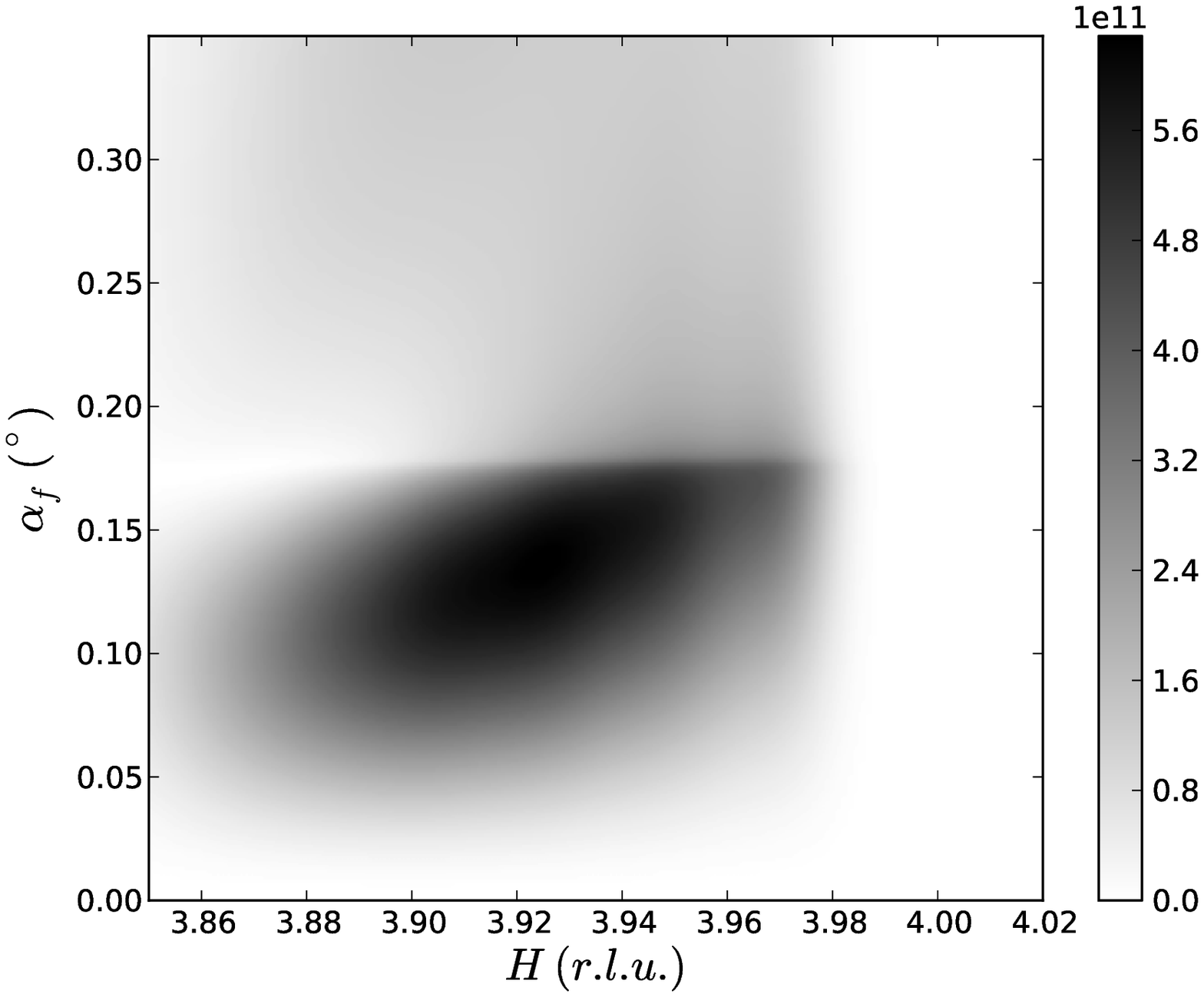}
\end{figure}

In Fig.\ref{figDWBA} is shown the simulated scattering for a germanium quantum dot on a silicon substrate. For the sake of demonstrating the use of PyNX, a simple analytical model is used:
\begin{itemize}
\item the quantum dot shape corresponds to a portion of a sphere, with a radius equal to 50 unit cells and a height of 20 unit cells.
\item the germanium content varies linearly from $20\%$ (bottom) to $80\%$ (top)
\item the relaxation (x,y,z being the fractional coordinates relative to the silicon substrate lattice) are:\\ $\varepsilon_{xx}=\varepsilon_{yy}=0.005+z*0.002*(1+\frac{1}{50}\sqrt{x^2+y^2}) $
\end{itemize}

Moreover, in this example only the scattering from the quantum dot is calculated, ignoring any contribution from the substrate. The corresponding script is provided as a supplementary file. A more complete description of the scattering, taking into account scattering from the (strained) substrate \cite{schmidbauer_effects_2005}, could also be added in the future.

\section{Availability}

The PyNX library is freely available from the project website at: http://pynx.sourceforge.net. It is open-source software, distributed under the CeCILL-B License (http://www.cecill.info/index.en.html), a permissive license similar to the FreeBSD license.

Although this library has been developed and tested only under linux, it should work on any operating system (including MacOS X and Windows) supported by the  PyCUDA \cite{klaeckner_pycuda:_2009} library.

It has been tested on a variety of graphics card (9600 GT, GTX 295, 8800 GTX). Although it is recommended to use a dedicated graphics card (not used for display) for GPU computing, it is not a requirement - the library automatically cuts the number of atoms in order to decompose the calculation in batches which last less than 5s (a limit imposed by the CUDA library for graphics cards attached to a display). And it is also possible to use the CPU for calculations.

This library uses the scipy (http://www.scipy.org) and PyCUDA \cite{klaeckner_pycuda:_2009} libraries, and optionally the cctbx \cite{grosse-kunstleve_computational_2002} to determine the refraction index for the computing of transmission and reflection coefficients for grazing incidence X-ray scattering.

\section{Conclusion }\label{conclusion}

The main interest from this computing project is the ability to compute scattering for non-ideal structures without any approximation. This is particularly important for strained nano-structures where the calculation often uses a fast Fourier transform approximation, even though the displacements from the ideal structure are large. This could also be useful for coherent diffraction imaging in Bragg condition for strained nano-structures \cite{minkevich_inversion_2007,robinson_coherent_2009,minkevich_selective_2009,favre-nicolin_analysis_2010,diaz_imaging_2010}, especially in order to extend this method to severely distorted lattices (e.g. near an epitaxial interface with dislocations, a grain boundary,...).

A current limitation of this project is related to the toolkit used - the CUDA development package is the most popular GPU computing tool available at the moment, but it depends on a single manufacturer, and remains proprietary. An important development in that regard is the creation of the OpenCL language (http://www.khronos.org/opencl/), which is intended to allow GPU-computing \textit{independently of the graphics card}. A future implementation of the proposed algorithm could use OpenCL and ensure its usability on a larger range of computing equipment.

\ack{The authors would like to thank Thierry Deutsch and Fr\'ed\'eric Lan\c con for helpful discussions during the development of this software package.}

\referencelist[GPU]

\end{document}